%% file: main.tex
\documentclass[chi_draft]{sigchi}

\makeatletter
\def\@copyrightspace{\relax}
\makeatother

\CopyrightYear{2020}
\setcopyright{acmlicensed}
\doi{https://doi.org/10.1145/3313831.XXXXXXX}
\isbn{978-1-4503-6708-0/20/04}
\conferenceinfo{UIST'20,}{October  20--23, 2020, Minneapolis, MN, USA}
\acmPrice{\$15.00}

\usepackage[utf8]{inputenc} 
\usepackage{balance}       
\usepackage{graphics}      
\usepackage[T1]{fontenc}   
\usepackage{txfonts}
\usepackage{mathptmx}
\usepackage[pdflang={en-US},pdftex]{hyperref}
\usepackage{color}
\usepackage{booktabs}
\usepackage{textcomp}
\usepackage{listings}
\usepackage{xspace}
\usepackage{cuted}
\usepackage{capt-of}

\definecolor{dkgreen}{rgb}{0,0.6,0}
\definecolor{gray}{rgb}{0.5,0.5,0.5}
\definecolor{mauve}{rgb}{0.58,0,0.82}
\definecolor{darkgray}{rgb}{0.33,0.33,0.33}

\newcounter{exampleline}[table]
\newcommand{\exampleline}[0]{\refstepcounter{exampleline}\theexampleline}

\lstdefinelanguage{ThingTalk}{
  keywords={class, extends, skill, let, function, notify, aggregate, of, new},
  keywordstyle=\color{blue}\bfseries,
  ndkeywords={this, query, list, sum, avg, timer, attimer},
  ndkeywordstyle=\color{darkgray}\bfseries,
  identifierstyle=\color{black},
  sensitive=false,
  comment=[l]{//},
  morecomment=[s]{/*}{*/},
  commentstyle=\color{purple}\ttfamily,
  stringstyle=\color{red}\ttfamily,
  morestring=[b]',
  morestring=[b]"
}

\usepackage{microtype}        
\usepackage{ccicons}          

\newcommand{\TODO}[1]{{\color{blue} TODO: #1}}

\def\plaintitle{VASH: Multi-Modal Programming of Web-Based Virtual Assistant Skills}
\def\emptyauthor{}
\def\plainkeywords{}

\makeatletter
\def\url@leostyle{%
  \@ifundefined{selectfont}{
    \def\UrlFont{\sf}
  }{
    \def\UrlFont{\small\bf\ttfamily}
  }}
\makeatother
\def\pprw{8.5in}
\def\pprh{11in}

\definecolor{linkColor}{RGB}{6,125,233}
\hypersetup{%
  pdftitle={\plaintitle},
  pdfauthor={\emptyauthor},
  pdfkeywords={\plainkeywords},
  pdfdisplaydoctitle=true, 
  bookmarksnumbered,
  pdfstartview={FitH},
  colorlinks,
  citecolor=black,
  filecolor=black,
  linkcolor=black,
  urlcolor=linkColor,
  breaklinks=true,
  hypertexnames=false
}

\newcommand{\WebTalk}[0]{WebTalk\xspace}

\newcommand{\Webber}[0]{VASH\xspace}
\newcommand{\VASH}[0]{VASH\xspace}

\newcommand{\Webops}[0]{Web primitives\xspace}
\newcommand{\webops}[0]{web primitives\xspace}
\newcommand{\Webop}[0]{Web primitive\xspace}
\newcommand{\webop}[0]{web primitive\xspace}

\begin{document}

\title{Multi-Modal End-User Programming of \\
Web-Based Virtual Assistant Skills}

\numberofauthors{1}
\author{%
  \alignauthor{Michael H. Fischer \quad Giovanni Campagna \quad Euirim Choi \quad Monica S. Lam\\
    \affaddr{Computer Science Department}\\
    \affaddr{Stanford University}\\
    \email{\{mfischer, gcampagna, euirim, lam\}@cs.stanford.edu}\\
    }
  }



\maketitle

\begin{strip}\centering
\includegraphics[width=\linewidth,keepaspectratio]{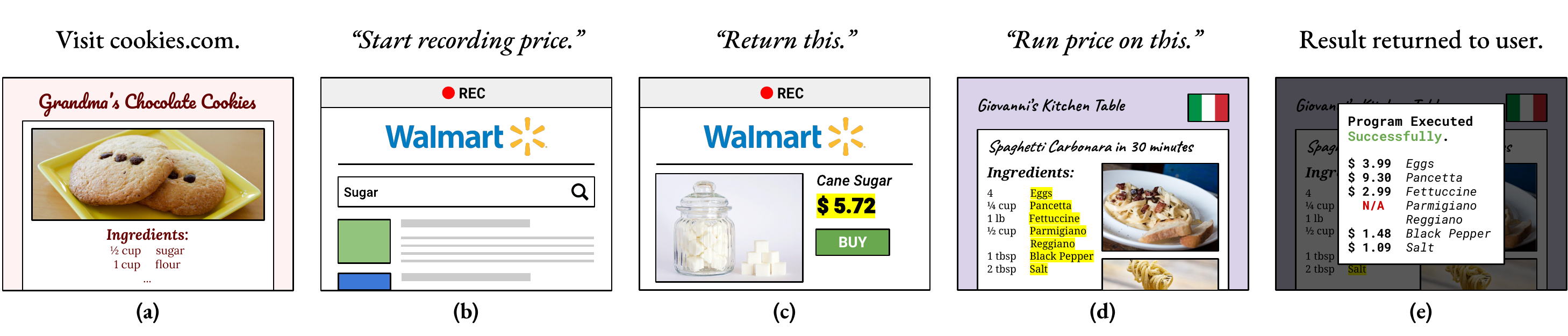}
\captionof{figure}{A scenario describing the process to create a program by demonstration using VASH.
(a) A user sees a cookie recipe on a popular food blog and wants to see how much the ingredients are. (b) He then enters VASH's recording mode using his voice and searches for one of the ingredients on Walmart's website. (c) He clicks on the first search result and highlights the price, telling VASH via voice that it should be returned. (d) A few weeks later, he is interested in the ``Spaghetti Carbonara" recipe on another food blog. He highlights the ingredients and asks VASH to run the previously defined program with them. (e) VASH returns to the prices of the items, but also knows to notify him that one of the ingredients is not available on Walmart.com.
\label{fig:feature-graphic}}
\end{strip}

\begin{abstract}
\input{abstract}

\end{abstract}



\input{introduction}

\input{related_work}
\input{overview}
\input{lang}
\input{algo}
\input{implementation}

\input{user_study}
\input{limitations}
\input{conclusion}

\balance{}

\bibliographystyle{SIGCHI-Reference-Format}
\bibliography{sample}

\end{document}

%% file: abstract.tex
 While Alexa can perform over 100,000 skills on paper, its capability covers only a fraction of what is possible on the web. To reach the full potential of an assistant, it is desirable that individuals can create skills to automate their personal web browsing routines. Many seemingly simple routines, however, such as monitoring COVID-19 stats for their hometown, detecting changes in their child’s grades online, or sending personally-addressed messages to a group, cannot be automated without conventional programming concepts such as conditional and iterative evaluation. This paper presents VASH (Voice Assistant Scripting Helper), a new system that empowers users to create useful web-based virtual assistant skills without learning a formal programming language. 

With VASH, the user demonstrates their task of interest in the browser and issues a few voice commands, such as naming the skills and adding conditions on the action. VASH turns these multi-modal specifications into skills that can be invoked in voice on a virtual assistant. These skills are represented in a formal programming language we designed called WebTalk, which supports parameterization, function invocation, conditionals, and iterative execution. 

\VASH is a fully working prototype that works on the Chrome browser on real-world websites.  
Our user study shows that users have many web routines they wish to automate, 81\% of which can be expressed using VASH.  We found that \VASH is easy to learn, and that a majority of the users in our study want to use our system.

%% file: introduction.tex
\section{Introduction}
Virtual assistants provide consumers with a new interaction paradigm: instead of interacting graphically with a web browser, users can now issue a wide variety of voice commands, from turning on their lights to controlling their music. 
Today, Alexa has an open skill platform, with over 100,000 service providers having implemented skills by defining how voice commands map to invocations of their APIs~\cite{voicebot2019}. While having 100,000 skills is impressive, its capability pales in comparison to what can be done on the web across over 1.8B websites. Furthermore, skills require the existence of APIs that smaller websites often don't have. Could we make a virtual assistant that can work beyond APIs and operate across the web? This paper explores how end users, even those without programming experience, can leverage the web to construct new virtual assistant skills for personal use.

Figure~\ref{fig:feature-graphic} shows an example of a personalized skill that a typical user may want to build. The skill takes a recipe, finds the prices of all the ingredients, and shows their sum to user. It combines two different websites, making it unlikely a dedicated API combining both exists. It involves performing operations such as iteration and aggregation, which current virtual assistants are unable to do. Yet, it is something that the user is already capable of performing by hand on the web. Our need-finding survey shows that there are many such tasks that users are performing repeatedly on the web that they would like to automate. 

Our system, \VASH (Virtual Assistant Scripting Helper), allows the user to automate web-based tasks by converting them into personalized virtual assistant skills. \VASH features a novel multi-modal interface combining Programming By Demonstration (PBD)~\cite{10.5555/168080} and voice commands. In PBD, the interaction of the user with the web page is recorded and converted into executable operations. The PBD operations and the voice commands are converted to a formal \WebTalk language, which defines the capabilities of the system, and provides compositionality to arbitrary programs. \WebTalk includes high-level programming concepts, such as function definitions, parameter passing, iteration, and filtering. 

There are a number of challenges in building a system that supports end-user programming on the web in a manner that is both useful and usable.

\subsection{Expressiveness: control constructs}
Many of the tasks that users want require performing actions across domains, such as finding the best rated restaurant on Yelp, looking up reservations on Opentable, or checking their bank balance prior to making a payment. There are many possible combinations of functions, and every user has their own routine they wish to automate. 

Almond is a virtual assistant that lets users perform compound commands that involve multiple skills. It proposes using a formal virtual assistant programming language, called ThingTalk, as the target language for natural language translation via a neural semantic parser. ThingTalk is designed to be translatable from natural language; as such, it has a very simple control construct. A ThingTalk program can connect together at most an event, data retrieval operations and one action, along with filters on the parameter and result values. Each of the event, data retrieval, and action are pre-defined primitive functions, which involves calling remote APIs using JavaScript. This language is not expressive enough for the kind of tasks we wish to automate.  

Given that we need to introduce concepts of control constructs in user-defined tasks, it is even more important that we use a formal programming language with well-defined semantics as a target language. Thus we ask the question: 
{\em How do we improve the expressiveness of the virtual assistant programming language to cover the tasks we wish to automate?}  

We propose \WebTalk, as an extension to ThingTalk to (1) include user-defined function definition and invocation, iterations, and conditional execution, and (2) defining functions as web page operations. Even though such control flow is common in all programming languages, the unique consideration here is whether it is a good target for translation from an end-user's specification. 

\subsection{End-user programming: Multimodal programming}

To let individuals automate their personal tasks, they must be able to specify and execute them. A purely voice-based approach is infeasible because it is hard to describe operations on web pages. A purely graphical approach using programming by demonstration (PBD)~\cite{li2017sugilite} is also inadequate because it is hard to express the concepts of control constructs. Recently, PixelTone~\cite{laput2013pixeltone} proposed adding voice to PBD, but their work cannot handle programming constructs. 

{\em Would end-users be able to program these web-based skills, which may require sophisticated control constructs?}  We propose \VASH, a PBD system with a multi-modal interface that encompasses a browser extension and a conversation voice interface that supports control structures. 

Our system simplifies programming with control constructs, using input and output modalities that are well suited for each construct. For example, voice and natural language are well suited for specifying control flow. Keyboards are well suited for quickly allowing the user to specify text. The mouse is easy to use for specifying clicks and actions. Finally, audio from the computer is well suited for providing feedback from the system. 

\subsection{Contributions}
Our paper makes the following contributions:

 \begin{enumerate}
\item 
We conduct a need-finding survey and show that there is an interest and need for end-users to automate their digital tasks on the web. Many such tasks require specification of control structures iterations, conditionals, triggers, and function composition, many of which cannot be achieved using today's virtual assistants or PBD systems. 

\item 
We have developed the first multi-modal interface that allows end-users to specify web-based virtual assistant skills---which support the use of control structures---by demonstration and voice. This interface supports 81\% of the tasks we collected in our need-finding user survey.

\item 
To formalize the capability of the PBD system, we have developed \WebTalk, the first virtual assistant programming language that is sufficiently expressive to support operations with control structures on web pages. The language is high-level and can be used as a translation target from the multi-modal end-user specification. 

\item We show our design is easy to learn and use through a user study involving five tasks in a controlled environment, where users are exposed to increasingly more complex tasks.

\item We developed a full end-to-end prototype of the \VASH design, and in a user-test with four real-world scenarios we find 80\% of the users would find our prototype useful, and 53\% agree they would use it. 
\end{enumerate}

%% file: related_work.tex
\section{Related Work}

\subsection{In-Browser Virtual Assistants}

Several parties have explored the possibility of integrating virtual assistants, often with voice interfaces, into the browser. Hey Scout, a browser-based voice assistant~\cite{tsai_kaye_scout} that enabled users to do simple web browsing tasks via a natural language interface later paved the way for Mozilla's public release for Firefox of a similar system~\cite{mozilla_2020}. Unlike our system, however, neither of these assistants interact with web page content.

\subsection{Web Automation}

Other works have focused on the popular problem of web automation via programming by demonstration, enabling non-programmers to automate complex online tasks. Rousillon~\cite{rousillon_chasins}, focused on the narrow task of web scraping automation via PBD, assessing system usability in a user study with computer scientists. Ringer~\cite{ringer_barman} allowed users to automate a more general set of online tasks including filling forms, but did not enable program parametrization without delving back into programming.

\subsection{Multi-modal Interfaces}

With early PBD systems often being difficult to use, some systems have focused on improving usability using multi-modal interfaces. Using computer vision and language, VASTA~\cite{Sereshkeh_2020} enables program automation across operating systems or platforms by intelligently recognizing interactive elements on screens. SUGILITE~\cite{li2017sugilite} and APPINITE~\cite{li2018appinite} provide users multi-modal interfaces in the form of primitive dialogue systems and visual interfaces to facilitate smartphone automation. Brassau~\cite{fischer2018brassau} automatically generates graphical user interfaces based on a natural language command.  These systems, however, did not provide general support for the program control structure supported by our system. 

\subsection{Dialogue Systems For Automation}

Other systems have attempted to end-user web automation primarily using natural language. Almond~\cite{almondwww17}, one of the most prominent of such systems, is an open-source virtual assistant that lets users issue compound natural language commands that compose of two of the three clauses: "when" a condition happens, "get" some data, and "do" some action. Natural language sentences are translated into ThingTalk programs, which consist of a single statement that connect together up to two skills. Each skill maps to some API in the Thingpedia repository, and has to be created by programmers, limiting the scope considerably. ThingTalk also does not support high-level programming concepts such as function definitions.

\subsection{Introductory Programming Systems}

It can be difficult for non-programmers to understand programming concepts, as the way in which non-programmers solve challenges can often be different from their programmer counterparts~\cite{grover2013computational, pane2001studying}. In this work we build off previous work in understanding the mental models of introductory programming education in information visualization for novices~\cite{burnett2003end}, practical trigger-action programming~\cite{ur2014practical}, and end user uses of spreadsheets~\cite{burnett2003end}.

%% file: overview.tex
\begin{table*}[htb]

\small
\centering
\begin{tabular}{l|p{4cm}|p{9.8cm}}
\toprule
\bf \VASH Web Primitives       & \bf \WebTalk Web Primitives & \bf Description \\
\hline
Open page (\textit{url})  & $@\text{load}(\textit{url})$ & Navigate the browser to the given url.\\
\hline
Click (\textit{element})           & $@\text{click}(\textit{selector})$ & Click on the elements matching the CSS selector.
\\
\hline
Type (\textit{element, value})             & $@\text{set\_input}(\textit{selector}, \textit{value})$ & Set the input elements matching the CSS selector to the given value.\\
\hline
Select (\textit{element})          & $\textbf{let}~\text{this}~:=~@\text{select}(\textit{selector})$ & Read the text in each element matching the CSS selector. \\
\hline
Paste (\textit{element)}           & $@\text{set\_input}(\textit{selector}, \textit{value}=\text{this})$ & Set the input elements matching the CSS selector to the content of the current selection, or the implicit input parameter of the current function. \\
\bottomrule
\end{tabular}
\caption{\Webops that the user can perform in \VASH, and the corresponding \WebTalk statements.  ``CSS selector'' refers to the \textit{selector} derived from the \textit{element} used during demonstration.}
\label{table:webops}
\end{table*}

\begin{table*}[htb]
\small
\centering
\begin{tabular}{p{3.7cm}|p{6cm}|p{7.0cm}}
\toprule
\bf \VASH Constructs         & \bf \WebTalk Constructs & \bf Description \\
\hline
``Start recording <func-name>'' & $\textbf{function}~\text{<func-name>}() \{$ & Begin recording a new function, with the given name.\\
\hline
``Stop recording'' & $\}$ & Complete the current function definition and save it for later invocation.\\
\hline
``Run <func-name>'' & $\text{<func-name>}()$ & Execute a previously-defined function.\\
\hline
``Run <func-name> with this'' & $\text{this} \Rightarrow \text{<func-name>}(\text{this}.\text{text})$ & Iterate over all elements of the current selection, and execute the function with each element as the input parameter.\\
\hline
``Return this value'' & $\text{this} \Rightarrow \textbf{notify}$ & Use the current selection as the return value of the current function.\\
\hline
``Run <func-name> if <cond>'' & $\text{this}, \text{<cond>} \Rightarrow \text{<func-name>}()$ & Execute a previously-defined function if the condition on the current selection is satisfied.\\
\hline
``Run <func-name> with this \newline if <cond>'' & $\text{this}, \text{<cond>} \Rightarrow \text{<func-name>}(\text{this}.\text{text})$ & Iterate over all elements of the current selection, and execute the function with each element that satisfies the condition.\\
\hline
``Run <func-name> at <time>'' & $\text{timer}(\text{<time>}) \Rightarrow \text{<func-name>}()$ & Execute the function every day at the given time.\\
\hline
``Run <func-name> with this at <time>'' & $\text{timer}(\text{<time>}) \Rightarrow \text{this} \Rightarrow \text{<func-name>}(\text{this}.\text{text})$ & At the given time, iterate over all elements of the current selection, and execute the function with each element.\\
\hline
``Return this value if <cond>'' & $\text{this}, \text{<cond>} \Rightarrow \textbf{notify}$ & Use the current selection as the return value of the current function, if the condition is satisfied.\\
\hline
``Calculate the <agg-op> of this'' & $\textbf{let}~\text{<agg-op>} := \textbf{aggregate}~\text{<agg-op>}~\text{number}~\textbf{of}~\text{this}$ & Compute the given aggregation operator based on the numeric values in the currently selected elements, and save it as a variable.\\
\hline
``This is a <var-name>'' \newline(when an input box is focused) & $@\text{set\_input}(\textit{selector}, \textit{value}=\text{<var-name>})$ & Create a new function parameter with the given name, and use it to set the input elements matching the selector.\\
\hline
``This is a <var-name>'' \newline(when some element is selected) & $\textbf{let}~\text{<var-name>}~:=~@\text{select}(\textit{selector})$ & Create a new named variable and set the elements currently selected to it.\\
\hline
``Return the <var-name>'' & $\text{<var-name>} \Rightarrow \textbf{notify}$ & Use the given named variable as the return value of the current function.\\
\hline
``Run <func-name> with \newline <var-name>'' & $\text{<var-name>} \Rightarrow \text{<func-name>}(\text{this}.\text{text})$ & Iterate over all elements of the given named variable, and execute the function with each element as parameter.\\\hline
``Start selection'' & n/a & Enter selection mode, allowing the user to select multiple non-contiguous elements. \\
\hline
``Stop selection'' & n/a & Exit selection mode. \\
\bottomrule
\end{tabular}
\caption{Constructs that \VASH understands, and corresponding WebTalk statements. The user issues each construct as a voice utterance. The table includes only the canonical form of each utterance.}
\label{table:constructs}
\end{table*}

\begin{table*}
\small
\centering
\begin{tabular}{lllr}
\toprule
\multicolumn{2}{l}{\bf \VASH Specification}                & \multicolumn{2}{l}{\bf \WebTalk Code} \\
\midrule
\Webop: & Select and copy ``AAPL'' & 
$\textbf{let}~\text{this}~:=~@\text{select}(\textit{selector}=\text{``a.company:nth-child(3)''});$ & (\exampleline)\\
Construct: & ``Start recording stocks''  & $\textbf{function}~\text{stocks}(\textit{param} : \text{String})~\{$ & (\exampleline)\\
\Webop: & Open finance.yahoo.com &
$~~~@\text{load}(\textit{url}=\text{``https://finance.yahoo.com''});$ & (\exampleline)\\
\Webop: & Paste in the search box & $~~~@\text{set\_input}(\textit{selector}=\text{``input\#search''}, \textit{value}=\textit{param});$ & (\exampleline)\\
\Webop: & Click Search button &
$~~~@\text{click}(\textit{selector}=\text{``button[type=submit]''});$ & (\exampleline)\\
\Webop: & Select current price &
$~~~\textbf{let}~\text{this}~:=~@\text{select}(\textit{selector}=\text{``span\#today-quote''});$ & (\exampleline)\\
Construct:& ``Return this value'' &
$~~~\text{this}~\Rightarrow~\textbf{notify};$ & (\exampleline)\\
Construct:& ``Stop recording'' &
$\}$ & (\exampleline)\\
\\
\Webop: & Select a stock symbol & 
$\textbf{let}~\text{this}~:=~@\text{select}(\textit{selector}=\text{``span.symbol:nth-child(1)''});$ & (9)\\
Construct: & ``Run stock with this'' & 
$\text{this} \Rightarrow \text{stock}(\text{this}.\text{text});$ & (10)\\
\bottomrule
\end{tabular}
\caption{A Basic Example: Get stock quote. The user performs the actions in the left column, and the corresponding \WebTalk program in the right column is generated. The value of the \textit{selector} parameter is a CSS selector that refers to a specific HTML element, and is generated from the element selected, clicked or written into.
CSS selectors are simplified in the example for illustration purposes.}
\label{table:basic_example}
\end{table*}
\section{Overview}

\VASH is a system that lets users create skills consisting of operations involving web page(s) from one or more websites. These skills are entered into the user's personal skill repository, which can be later invoked by voice. 

\VASH lets users program by demonstration (PBD). The user can behave as they normally do without learning how to articulate what they wish to automate. This is particularly important for GUI-based computations. 

However, when we show a computation by demonstration, we only work with {\em concrete} values, or actual values, which result in concrete results for those inputs. On the other hand, a function needs to be parameterized, with the results dependent on the input values. For example, suppose we wish to find the best Italian restaurant on a web page. As users, we would find the first Italian restaurant on a list sorted by the ratings. Just recording the click on a particular Italian restaurant fails to capture the reason behind the action.  It is impossible to intuit how to generalize skills to new inputs without users' help. \VASH lets the users use voice to describe all additional information to turn the demonstration into a program that works across different inputs.

Unbeknownst to the users, underlying \VASH is a formal programming language, called WebTalk, that we created for this purpose. WebTalk is designed to be translated from multimodal inputs, with voice, the mouse, and the keyboard. The language supports variables, function calls (without recursion), iterations, and conditional executions. Grounding the PBD system with a formal language provides consistency and compositionality. Our goal is to give users a clean, implicit conceptual model so they can come to know what to expect, instead of confronting them with a set of ad hoc features. This allows them to become more proficient over time. While we do not expect users to create long function bodies, they can easily define functions that call other functions, allowing their capability to compound. 

\VASH is a programming system that accepts a multi-modal specification via PBD, translates it into \WebTalk program, and runs it. The specification of a skill has two modalities: 

\begin{enumerate}
\item The web browsing actions using the mouse and keyboard; these are translated into {\em web primitives} in WebTalk. 

\item Voice commands to provide meta-information that translates the demonstration into parameterized functions; these are translated into {\em constructs} in WebTalk. 
\end{enumerate}

%% file: lang.tex
\section{\VASH Specification and WebTalk}
\VASH and \WebTalk are designed hand-in-hand to be easy to use for consumers to automate their routines on the web, trading away functionality where necessary. 
 
The correspondence between the \VASH specification and the WebTalk language is shown in Tables~\ref{table:webops} and ~\ref{table:constructs}. Note that users do not have to use exactly the phrases listed in the \VASH specification, as \VASH uses natural language understanding to interpret the statements.
Due to space, we will not detail the WebTalk grammar, except to note a few less common conventions, which are borrowed from ThingTalk, a virtual assistant programming language~\cite{almondwww17}. Variables are tables of values; a scalar is a degenerate table with one row. A statement of the form:
\[\text{<var-name>}, \text{<cond>} \Rightarrow \textit{op}\]
says ``perform $\text{op}$ for every value in the table $\text{<var-name>}$ satisfying the condition $\text{<cond>}$''. The condition will name a field in the table. Tables in \WebTalk correspond to HTML elements, and have fields \textit{text} (the element's textual content) and \textit{number} (the element's numeric value, if the element contains a number).
The keyword $\textbf{notify}$ returns the result; if the command is invoked by the user directly, the result is shown to the user in a popup.


\subsection{Basic Function Definitions}

We first describe a use case as a running example to illustrate the basic features of the \VASH system. Ann is reading a story about Apple, when she realizes she keeps looking up stock quotes for companies in stories she reads. Using \VASH, she creates a skill for herself  to return the quote of a stock, as shown in Table~\ref{table:basic_example}. The first column details what Ann does, classified into GUI actions which generate web primitives in WebTalk, and voice commands which generate WebTalk constructs. As shown by the sequence of web primitives in this example, she looks up the stock quote on finance.yahoo.com as she normally would. She adds some voice commands to turn the actions into a function, which she invokes with possibly a different stock symbol afterwards. 

\subsubsection{\Webops}
Let us first discuss the GUI actions. 
The function we define needs to take the user's demonstration with a concrete input value and produce a parameterized version that works with new inputs. First, we must record all keyboard input in forms, mouse clicks on buttons and links, as well as select, copy and paste. We do not need to record operations such as scrolling or moving the mouse, as those operations only affect the view of the users. Drawing with the mouse, which involves a click and a drag, is not currently supported.

Second, we need to make the web primitives relative to the new input parameters.  For example, when Ann selects the stock price, we need to capture the HTML element selected, and not the current value of AAPL, because we wish to run this function on a different stock symbol and at a different time.  We represent the HTML element of interest with
a CSS (Cascading Style Sheets) selector~\cite{cssselectors}. CSS selectors are a language for describing a subset of HTML elements in a page, originally designed for styling. CSS selectors use semantic information to identify the elements (HTML tag name, author-specified ID and class on each element), positional and structural information (ordinal position in document order, parent-child relationship), and content information.

When recording the action, \VASH records which element the user is interacting with, and generates a CSS selector that identifies that element uniquely. When available, \VASH uses ID and class information to construct the selector, falling back to positional selectors when those identifiers are insufficient to uniquely identify the element. As such, the CSS selectors \VASH generate are robust to changes in the content of the page and small changes in layout. \VASH selectors work best with static HTML pages where IDs and classes are assigned according to the semantic meaning of each element.

\VASH records all selection operations, such as when Ann selects and then copies the stock name at the beginning of the recording (Table~\ref{table:basic_example} line 1). The selection is converted to a ``select'' web primitive (Table~\ref{table:webops}), which refers to the element that was selected, rather than recording the specific values.

The result of the ``select'' web primitive is bound to the special ``this'' variable, which then can be used in the subsequent web primitives (Table~\ref{table:basic_example} line 1, lines 6 and 7, and lines 9 and 10). ``this'' is a special variable that is implicitly defined and accessed while recording. During execution, it contains the value of the elements from the page at execution time. The ``this'' variable is a list of HTML elements, and it records the \textit{text} content and the \textit{number} value.

Paste operations are mapped to a ``set\_input'' \webop (Table~\ref{table:webops}). If the paste refers to something that was copied in the same function, the value is set to the text of the currently selected element (the current ``this'' variable). If instead, the paste operation in the input box is paired to a copy operation outside the function, the value is set to the implicit function parameter ``param'' (Table~\ref{table:basic_example} line 4). 
Finally, if the copied content comes from outside the browser, the paste operation generates a ``set\_input'' web primitive having value set to the constant value being pasted, as if the user typed it explicitly (Table~\ref{table:webops}). This design avoids additional copy-paste web primitives while maintaining both the natural flow of copy-pasting during recording, and the ability of the make the execution depend on the actual page it is executed on.

\subsubsection{Function Definitions}
With PBD, functions are defined as they are executed on a specific information.  In this example, Ann demonstrates the function with the input parameter by copying ``AAPL''. The function is defined by wrapping the actions with a ``Start recording <func\_name>'' and a ``Stop recording''. The parameter ``param'' is automatically defined if a copied value is used in the function. 
The user can refer to the ``this'' variable in their voice command, as shown in the ``Return this value'' or ``Run stock with this'' commands.

The user can also say ``Run stock with this at 10am every day''. The assistant runs the skill in the background automatically, then notifies the users of the results. This functionality is identical to that of the existing Almond assistant~\cite{almondwww17}.

At most one return statement can appear in the function, but the return statement need not be the last. It can be followed by additional web primitives, which do not affect the return value. This allows the function to perform ``clean up'' actions, such as logging out, before returning the result.

\subsection{Advanced Constructs}

\input{complex_example}
To support what users want to automate on the web, \VASH supports function composition, iteration, conditional execution, and aggregation. Due to limited space, we will just overview these features by way of a more interesting use case. Suppose Bob is taking up baking, as many are doing these days. He finds a new recipe website with many interesting recipes, but he wishes to keep the cost of each experiment below a budget. Table~\ref{table:complex_example} shows how he can create a recipe skill that looks up a recipe, looks up each ingredient in that recipe in Walmart, then sums up the price. 

\subsubsection{Function Definitions in Functions}
As in the previous example, Bob starts recording the ``recipe'' function, pastes the recipe name in the search box, and clicks Search (lines 2 to 4). Now Bob sees the full list of ingredients. He needs to compute the price of each ingredient, checking from his favorite grocery chain.

Bob can check the price with a ``price'' function, but he has not defined one yet. Fortunately, \VASH lets users define another function inside a function, because the user would not know what helper functions are needed ahead of time, and the user would not have the context to use PBD to define the function. For simplicity, all functions have global scope, meaning that they can be invoked anywhere, and they cannot use any variables from the lexical scope of the function definition.
To define the ``price'' function, Bob copies the first ingredient, then starts recording the new function. In the new function, Bob pastes the ingredient name (copied from the recipe), searches the product, selects the price, returns it, and completes the ``price'' function (lines 7 to 13). 

\subsubsection{Function Call, Iteration and Conditional}
With the ``price'' function in hand, Bob can now apply it to all ingredients. He goes back to the recipe page, and selects all the ingredients in the list, then says ``run price with this'' (line 15). Because he has selected multiple elements, the selection is iterated and the ``price'' function is called on each element, returning a list of prices.  Bob is shown the list of prices computed immediately. 

This example illustrates an important complication in PBD systems. We need to execute the functions called in a function definition because the user is always operating with concrete values. The results need to be returned to allow the user continue with the demonstration.  If the called function returns a result, the result is displayed as a popup on top of the current page.

Iteration can also be made conditional, by adding a predicate (Table~\ref{table:constructs}, line 3). In this case, the function is only applied to the list elements that satisfy the predicate. For example, ``call \textit{alert} with this if this is greater than 98.6'' can be expressed as:
\begin{tabbing}
123\=\kill
\>$\text{this}, \textit{number} > 98.6 \Rightarrow \textit{alert}(\textit{param}=\text{this}.\text{text});$
\end{tabbing}

\textit{number} is a field of the currently selected HTML elements (in the ``this'' variable) and it is computed by extracting any numeric value in the elements. Our current system only supports a single predicate, which can use equality, inequality, or comparison between the current selection and a constant. As the natural language technology improves, we expect in the future to support arbitrary logical operators (and, or, not). 

\paragraph{Aggregation}
Having obtained the list of prices, Bob now needs to compute the sum (line 16). \WebTalk supports all the aggregation operations supported by ThingTalk with database extensions~\cite{xu2020schema2qa}. Once a set of numbers is selected in either a table or list, the \textit{sum}, \textit{count}, \textit{average}, \textit{max}, and \textit{min} can be computed and returned, by issuing the voice command ``calculate the <operation> on this''.

The result of the aggregation is immediately displayed to Bob, so he can decide what to do with it. He can now say ``return the sum'', to use the aggregation result as the result of the current function (line 17). He could also have used the ``sum'' in a subsequent conditional command.  

\paragraph{Named Variables}
In addition to the ``this'' variable that is implicitly defined and used with selection and copy-paste operations, users can also create named variables. This allows the user to manipulate multiple selections at once. This is necessary when, for example, the defined function needs more than one parameter.   
Named variables can be used in place of ``this'' in conditional, function invocation and return statements.

\subsection{Design Rationale Discussion}
\VASH is unconventional when compared to typical programming languages because it is intended for consumers, and it is used in a PBD setting. 
Most functions can be defined without users naming variables; a single ``this'' variable is used in the \WebTalk program, which is derived from the GUI actions recorded while the user demonstrates the program. We allow the specifications of functions to be nested, even though all functions have the same global scope. This way, the user can define helper functions as they discover the need for them during demonstration, and use them immediately in a function definition. These two features reduce the mental load on the user, as confirmed by our user study described later.  

Nested function specification is especially important because \WebTalk does not have scoping like most other languages, which are usually specified with ``begin'' and ``end'' or ``\{'' and ``\}''. Scoping is hard when the programming is not visual. Iterators and conditional statements can only be applied to functions.
Iterators are expressed as applying functions to elements of a collection, rather than ``do'' or ``for'' statements, which are more powerful but harder for non-programmers in a PBD setting. \VASH conditionals do not have an ``else'' clause. In PBD where users are operating with concrete values, the users can only perform actions that follow from conditions the concrete values satisfy. In the future, we can add ``else'' clauses by letting sophisticated users refine a defined function with additional demonstrations using alternate concrete values. 

%% file: complex_example.tex
\begin{table*}
\small
\centering
\begin{tabular}{lllr}
\toprule
\multicolumn{2}{l}{\bf \VASH Specification}                & \multicolumn{2}{l}{\bf \WebTalk Code} \\
\midrule
Construct: & ``Start recording recipe''  & $\textbf{function}~\text{recipe}(\textit{param} : \text{String})~\{$ & (\exampleline)\\
\Webop: & Open allrecipes.com &
$~~~@\text{load}(\textit{url}=\text{``https://allrecipes.com''});$ & (\exampleline)\\
\Webop: & Paste in search box &
$~~~@\text{set\_input}(\textit{selector}=\text{``input\#search''}, \textit{value}=\textit{param});$ & (\exampleline\label{exampleline:complex_example:set_input_1})\\
\Webop: & Click Search button &
$~~~@\text{click}(\textit{selector}=\text{``button[type=submit]''});$ & (\exampleline)\\
\Webop: & Click the first result &
$~~~@\text{click}(\textit{selector}=\text{``.recipe:nth-child(1)''});$ & (\exampleline)\\
\Webop: & Copy the first ingredient & $~~~\textbf{let}~\text{this} := @\text{select}(\textit{selector}=\text{``.ingredient:nth-child(1)''});$ & (\exampleline)\\
Construct: & ``Start recording price'' &
$~~~\textbf{function}~\text{price}(\textit{param} : \text{String})~\{$ & (\exampleline)\\
\Webop: & Open walmart.com &
$~~~~~~@\text{load}(\textit{url}=\text{``https://walmart.com''});$ & (\exampleline)\\
\Webop: & Paste in search box &
$~~~~~~@\text{set\_input}(\textit{selector}=\text{``input\#search''}, \textit{value}=\textit{param});$ & (\exampleline)\\
\Webop: & Click Search button &
$~~~~~~@\text{click}(\textit{selector}=\text{``button[type=submit]''});$ & (\exampleline)\\
\Webop: & Select price of top result &
$~~~~~~\textbf{let}~\text{this}~:=~@\text{select}(\textit{selector}=\text{``.result:nth-child(1) .price''});$ & (\exampleline)\\
Construct:& ``Return this value'' &
$~~~~~~\text{this}~\Rightarrow~\textbf{notify};$ & (\exampleline)\\
Construct:& ``Stop recording'' &
$~~~\}$ & (\exampleline)\\
\Webop: & Select all ingredients & $~~~\textbf{let}~\text{this} := @\text{select}(\textit{selector}=\text{``.ingredient''});$ & (\exampleline)\\
Construct:& ``Run price with this'' &
$~~~\textbf{let}~\text{this} := \text{this} \Rightarrow \text{price}(\text{this}.\text{text});$ & (\exampleline\label{exampleline:complex_example:set_this_result})\\
Construct:& ``Calculate the sum of this'' &
$~~~\textbf{let}~\textit{sum} := \textbf{aggregate}~\text{sum}~\textit{number}~\textbf{of}~\text{this};$ & (\exampleline\label{exampleline:complex_example:calculate_sum})\\
Construct:& ``Return the sum'' &
$~~~\textit{sum}~\Rightarrow~\textbf{notify};$ & (\exampleline\label{exampleline:complex_example:return_sum})\\
Construct:& ``Stop recording'' &
$\}$ & (\exampleline)\\
\bottomrule
\end{tabular}
\caption{Sum The Price of Recipe Ingredients. The user performs the actions in left column, and the corresponding \WebTalk program in the right column is generated. CSS selectors are simplified in the example for illustration purposes.}
\label{table:complex_example}
\end{table*}

%% file: algo.tex
\section{The \VASH System}

\begin{figure}[htb]
  \centering
  \includegraphics[width=0.7\linewidth]{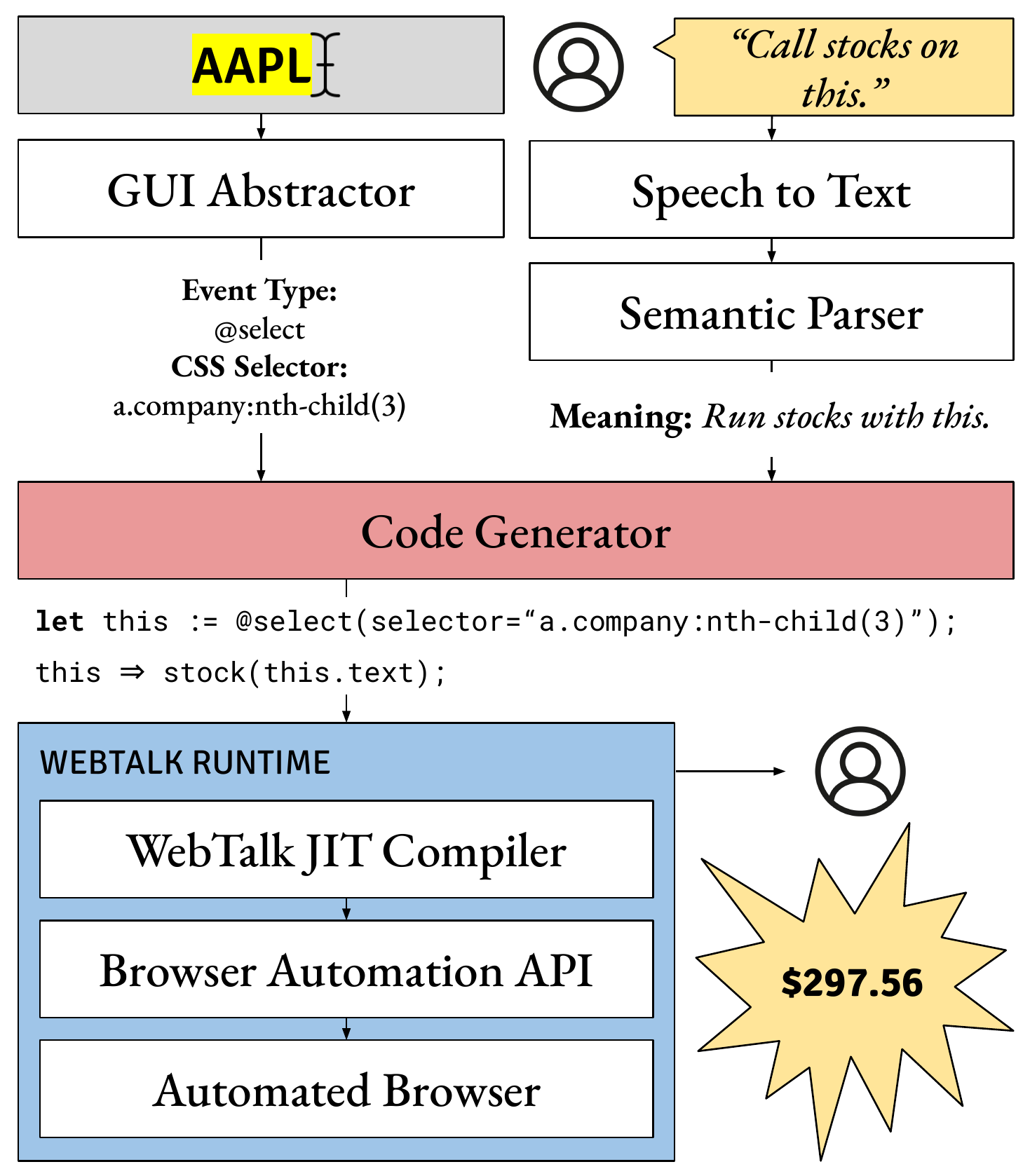}
  \caption{A high-level overview of the algorithm that \VASH uses to convert a multi-modal specification to \WebTalk.}
  \label{fig:arch}
\end{figure}

The \VASH system accepts a user's demonstration on the browser annotated with a small number of voice commands and translates the specification into 
\WebTalk. As discussed above, to support function calls in a function definition, the translator also includes a run-time system to execute \WebTalk programs. 
The high-level system architecture of \VASH is shown in Fig~\ref{fig:arch}.  


\subsection{GUI Abstraction}
The first step in converting the specification is to abstract all GUI operations that the user performs into \WebTalk \webops. To record all actions that the user performs on a page, \VASH uses a browser extension, which injects JavaScript on each page where \VASH is enabled. \VASH clearly displays to the user a prominent indicator when it is recording the user's actions. 

When the user starts recording a function, \VASH must first record the current context in which the function is recorded. \VASH records the current URL, and generates the corresponding @load \webop.

\VASH's injected JavaScript code listens to all interaction \textit{events} (keyboard, mouse, clipboard) from the browser on all parts of the page. When an event is intercepted, the injected JavaScript code considers the HTML element that is the \textit{target} of the event, and constructs a new CSS selector that identifies that element uniquely. The selector is constructed based on the ID, tag name and class name of the element. The event is then converted to the corresponding \webop.

To generate @set\_input, \VASH records keyboard events in input elements. It uses the value of the input element as the value of the \webop. Multiple consecutive keyboard events are collapsed into a single \webop: this allows the system to replace the \webop that contains a constant value with one that refers to a parameter when the user names that parameter.

\subsection{Natural Language Commands}
\VASH continuously listens for the user's voice, and reacts to the commands that map to \WebTalk control constructs (Table~\ref{table:constructs}). Each utterance is passed to a speech-to-text converter, followed by a semantic parser that translates it into a canonical \VASH command. VASH responds to the user's prompt by executing the command, and then replies to the user in voice.

\label{sec:selection-mode}
Commands that use ``this'' are interpreted according to the command issued immediately before, and the current selection on the page.
If the user just issued a ``run'' command that returns a result, ``this'' refers to the result of the called function. Else, if the currently focused element is an input element, ``this'' refers to the content of the input element, which is either a constant (what the user just typed) or a function parameter (that the user pasted or named explicitly). In that case, \VASH uses the recorded value of the input.
Otherwise, \VASH generates a @select \webop, using a CSS selector that maps to the currently selected elements. In the last case, ``this'' in natural language refers to the ``this'' variable in \WebTalk.

To allow the user to select complex lists of elements, and elements in tables or in pages with complex layouts, in addition to the plain browser selection, \VASH also supports an explicit \textit{selection mode}. The user enters selection mode with the voice command ``start selection''. While in selection mode, the page is not interactive: instead, clicks add or remove the clicked elements to the current selection.
Selection mode is exited with ``stop selection''. Selection mode is treated equivalently to a native browser selection operation, and generates a \WebTalk @select primitive.

\subsection{The Code Generator}
Given the stream of web primitives produced by the GUI abstractor, and the natural language commands, there is a straightforward translation to the corresponding \WebTalk code.

One exception is the ``run'' command: to allow the user to demonstrate a function call, when the user issues a ``run'' command while recording a function, \VASH must execute it as in the context of the page being recorded. To do so, \VASH constructs a \WebTalk program containing all the statements recorded so far, and the function call statement. In the generated program, the parameters of the current function are replaced with the concrete values they had when the corresponding \webops were recorded. When executed, the statements in the current functions will reestablish the context of the page so the ``run'' command can be executed faithfully. The result is then shown to the user in a popup. If the user issues multiple ``run'' commands in the same function, each time the whole current function will be executed from the beginning. This allows a clean separation of code generation from the \WebTalk runtime, at the expense of marginal inefficiency.

To support nested function definitions, \VASH maintains the full state of the code generator in a stack. When the user starts recording a new function, a new state is pushed on the stack, and it is popped when the recording is complete.

\subsection{\WebTalk Run-Time}

The \WebTalk code is first compiled to native JavaScript code, using a \WebTalk compiler. The compiler is based on the ThingTalk compiler~\cite{almondwww17}, extended to support nested functions. The JavaScript code is run on an \textit{automated browser}: a form of browser that is driven with an automated API rather than interactively by the user.
Each \WebTalk \webop is converted to a call to the automation API. Each \WebTalk function call is mapped to a new automation session (roughly corresponding to a new browser tab); this ensures that function calls do not interfere with the state of their parent. Other \WebTalk constructs (iteration, conditionals, aggregation) are implemented in the generated JavaScript code.


%% file: implementation.tex
\section{Implementation}

We implemented a fully functional end-to-end prototype for \VASH, written in JavaScript. The implementation is split in a Google Chrome browser extension that injects the \VASH recording code in every page the user visits, and a standalone Nodejs application containing the \WebTalk execution code. The standalone application is based on Almond~\cite{almondwww17}, and is able to spawn the automated browser and communicate with it. 

To handle the user's speech, we use the Web Speech API, a native speech-to-text and text-to-speech API available in Google Chrome. We  use the annyang library~\cite{annyang} to perform natural language understanding of the user's commands. This library uses a template-based NLU algorithm, requiring the user to speak exactly the supported words. At the same time, it supports open-domain understanding of arbitrary words, which is necessary to allow the user to define their own function names. Open-domain understanding is not reliable in currently existing neural NLU solutions, but we expect that future work will integrate a more flexible NLU model.

CSS Selectors are generated using the finder library~\cite{finder}. Event recording code is based off the Puppeteer Recorder Chrome Extension~\cite{puppeteer-recorder}, and event replaying uses the Puppeteer API~\cite{puppeteer} to automatically control Google Chrome.

It is not possible using browser APIs to distinguish between navigation explicitly caused by the user, and navigation triggered by clicking a link or submitting a form.  \VASH only allows one explicit navigation event per function currently.

%% file: user_study.tex
\section{Experimentation}
To evaluate our system, we performed four experiments: (1) We conduct a need-finding survey to learn what types of web flows users would like to automated. (2) We  evaluate whether users can learn the \VASH specification constructs. (3) We evaluate specific design choices in \VASH. (4) We collect user feedback on \VASH in user-suggested scenarios on real-world websites. 

\subsection{What Do Users Need To Automate?}
\begin{figure}
    \begin{center}
        \includegraphics[width=0.6\linewidth]{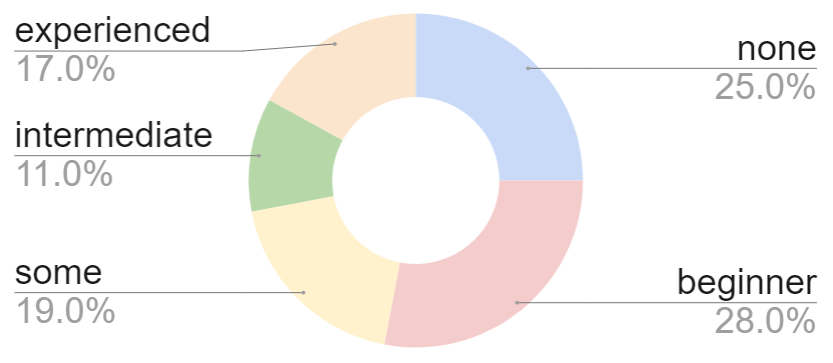}
    \end{center}
  \caption{Programming experience of survey participants that proposed skills for \VASH.}
  \label{fig:user_programming_1}
\end{figure}

\begin{figure}
    \begin{center}
        \includegraphics[width=0.6\linewidth]{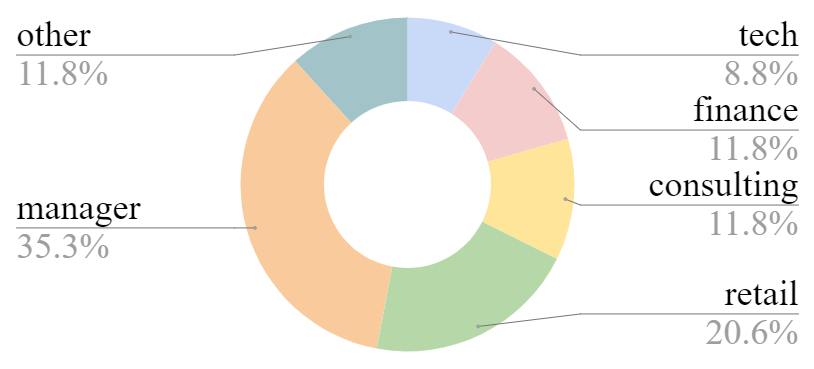}
    \end{center}
  \caption{Occupation of survey participants that proposed skills.}
  \label{fig:occupations_1}
\end{figure}

Our first study is a need-finding on-line survey to find out what users are interested in automating and whether the primitives in \VASH are adequate. We recruit 37 participants on Amazon Mechanical Turk (25 men and 12 women, average age = 34), each of whom was paid \$12 for approximately 60 minutes of their time. Survey participants had a mix of programming experience (Fig.~\ref{fig:user_programming_1}) and were from a variety of backgrounds (Fig.~\ref{fig:occupations_1}).


In the survey, respondents were first shown the functionality of the system, then asked to describe 3 skills each that they would like to be automated. We collected 71 valid skills. 

\begin{figure}
  \includegraphics[width=\linewidth]{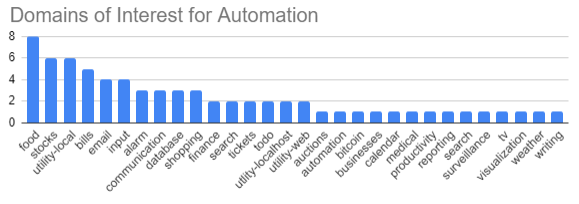}
  \caption{Domains of skills users were interested in creating using \VASH.}
  \label{fig:domains_of_interest}
\end{figure}

The proposed skills span 30 different domains, with the most popular being food, stocks, local utilities, and bills (Figure~\ref{fig:domains_of_interest}).
Representative tasks are shown in Table~\ref{tab:example_skills}. Of these 71 skills, we found
24\% do not require any programming construct,
28\% need iteration,
24\% need conditional statements,
24\% need a trigger (a timer plus a condition).
In summary, 76\% of the skills people want to automate require the control constructs we introduce to PBD.

\begin{figure*}
\includegraphics[width=\linewidth,trim={0 1cm 0 0}, clip]{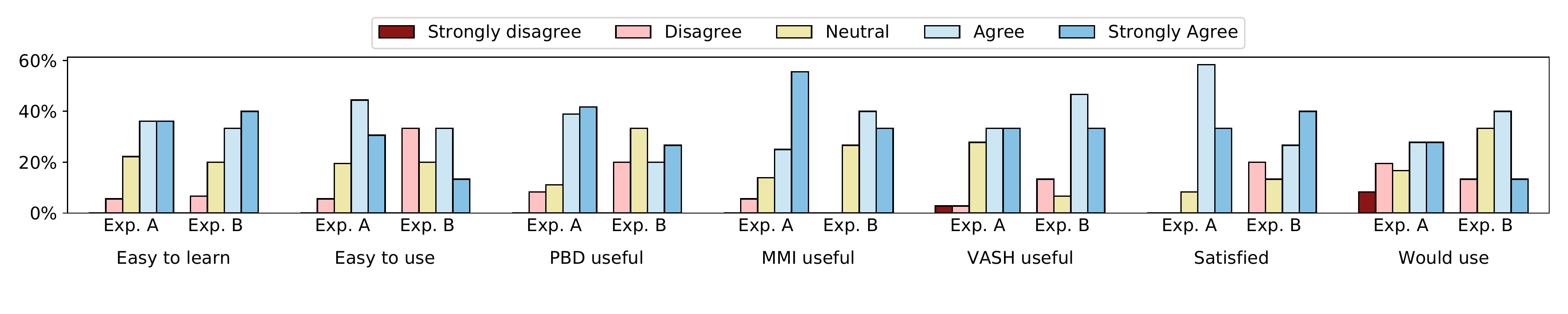}
\caption{Results of our user studies. ``Exp. A'' refers to the construct learning study, while ``Exp. B'' refers to the real world evaluation study.}
\label{figure:likert-scale}
\end{figure*}

 99\% of the skills are intended for the web and 1\% are to be run on the local computer. 34\% of skills are on websites that need authentication, showing that users are interested in skills that operate on their personal data. 
We found 81\% of the web skills can be expressed using \VASH. For the remaining 19\%, 11\% require producing charts, and 8\% require understanding videos and images. These functionalities are orthogonal and can be added to the system as pre-defined skills.  This shows that despite the simplicity of \VASH, it covers what people want to automate. 

\begin{table}[tb]
\small
\centering
\begin{tabular}{p{5.5cm}l}
\toprule
\bf Example Skill & \bf Domain \\
\midrule
``Make a reservation for the highest rated restaurants in my area.'' & Communication \\
\hline
``Order a ticket online if it goes under a certain price.'' &  Communication \\
\hline
``Check my investment accounts every morning and get a condensed report of which stocks went up and which went down.'' & Stocks \\
\hline
``Order ingredients online for a recipe I want to make, but only the ingredients I need.'' & Food \\
\hline
``Automate queries I do by hand every day for work for inventory levels and delivery times.'' & Database \\
\hline
``Send birthday text message to people automatically.'' &  Communication \\
\hline
``Alert me when someone moves on the camera of my home security system.'' & Unsupported \\
\bottomrule
\end{tabular}
\caption{Representative tasks that users wanted to automate.}
\label{tab:example_skills}
\end{table}

\subsection{Can Users Learn To Program In \VASH?}
Our next study addresses the question of whether users can effectively learn the programming constructs employed by \VASH. We conducted a remote user study, using the same participants as the need-finding survey. In the study, each participant was asked to perform a set of five tasks, with each task designed to be a realistic test of an element of the system’s control structure. The tasks were performed in the order of increasing complexity, to emulate the learning experience on the system.
The tasks were unsupervised and on demo websites that we had created for the purpose of testing the design of the system. For each task, users were asked to watch a video that demonstrated how the control structure construct worked so that they could understand the functionality of the system. After watching the video, they repeated the task. Lastly, they were asked to do a different task that requires the same construct. The five tasks they performed on their own are shown in Table~\ref{table:learnability-study}.
Note that because the ``Iteration'' task requires two parameters, the recipient name and their email address, the users have to name the parameters explicitly, instead of relying on the copied data as the implicit parameter. 

\begin{table}[t]
\small
\centering
\begin{tabular}{ll}
\toprule
{\bf Construct} & {\bf Task} \\
\midrule
Basic & Automated the clicking of a button. \\
Iteration & Send an email to a list of email addresses. \\
Conditional & Reserve a restaurant conditioned on rating. \\
Timer & Buy a stock at a certain time. \\
Filter & Show restaurants above a certain rating. \\
\bottomrule
\end{tabular}
\caption{Tasks performed by the participants in the programming construct study.}
\label{table:learnability-study}
\end{table}

\paragraph{Quantitative Results}
Participants were able to successfully complete the tasks in automating routines using \VASH 94\% of the time. After the tasks, users were asked to complete a survey, rating a number of questions on a 5-point Likert scale from ``strong disagre'' to ``strongly agree''. The results are shown in Figure~\ref{figure:likert-scale} as ``Exp. A''. We notice that users consistently found the system easy to learn (72\%), and easy to use (75\%). Both the PBD interface, and the multi-modal interface (``MMI'' in the plot) are rated helpful by 81\% of survey participants. Overall, 66\% of the users agree that \VASH is useful, and are 91\% are satisfied with the experience of testing it. Furthermore, 55\% of the users say they would use \VASH to automate their personal skills. These results confirm the need and usefulness of \VASH, and suggest that the programming constructs employed by \VASH can be learned. Note that in these experiments, users are only exposed to simple tasks on a custom website, rather than real world scenarios, which explains the low propensity to find \VASH useful.

\subsubsection{Privacy}
When automating a task that involves personal identifiable information, 83\% of user wanted a privacy preserving system that ran locally and 33\% did not. When automating any task, even ones that did not involve personal identifiable information, 66\% of users still wanted a privacy preserving system.  As our system is able to run on the client as a Chrome extension and native application, we are able to offer users privacy. 

\subsection{Evaluating Design Decisions}
Here we evaluate some of the decisions we made in the design of the \VASH specification, with a user study with 14 users  (7 men and 7 women, average age 25) conducted over video conferencing.

\subsubsection{Implicit Variables}
Instead of requiring users to define all their variables, \VASH introduces the implicit ``this'' variable that the users can define and use with select and paste actions. We ask the users to build an example skill using both the explicit and implicit naming methods. Overall, 88\% preferred the implicit 
version, because it had fewer steps and was faster. We found users didn't like talking to their computer as much.

\begin{figure}
 \includegraphics[width=\linewidth]{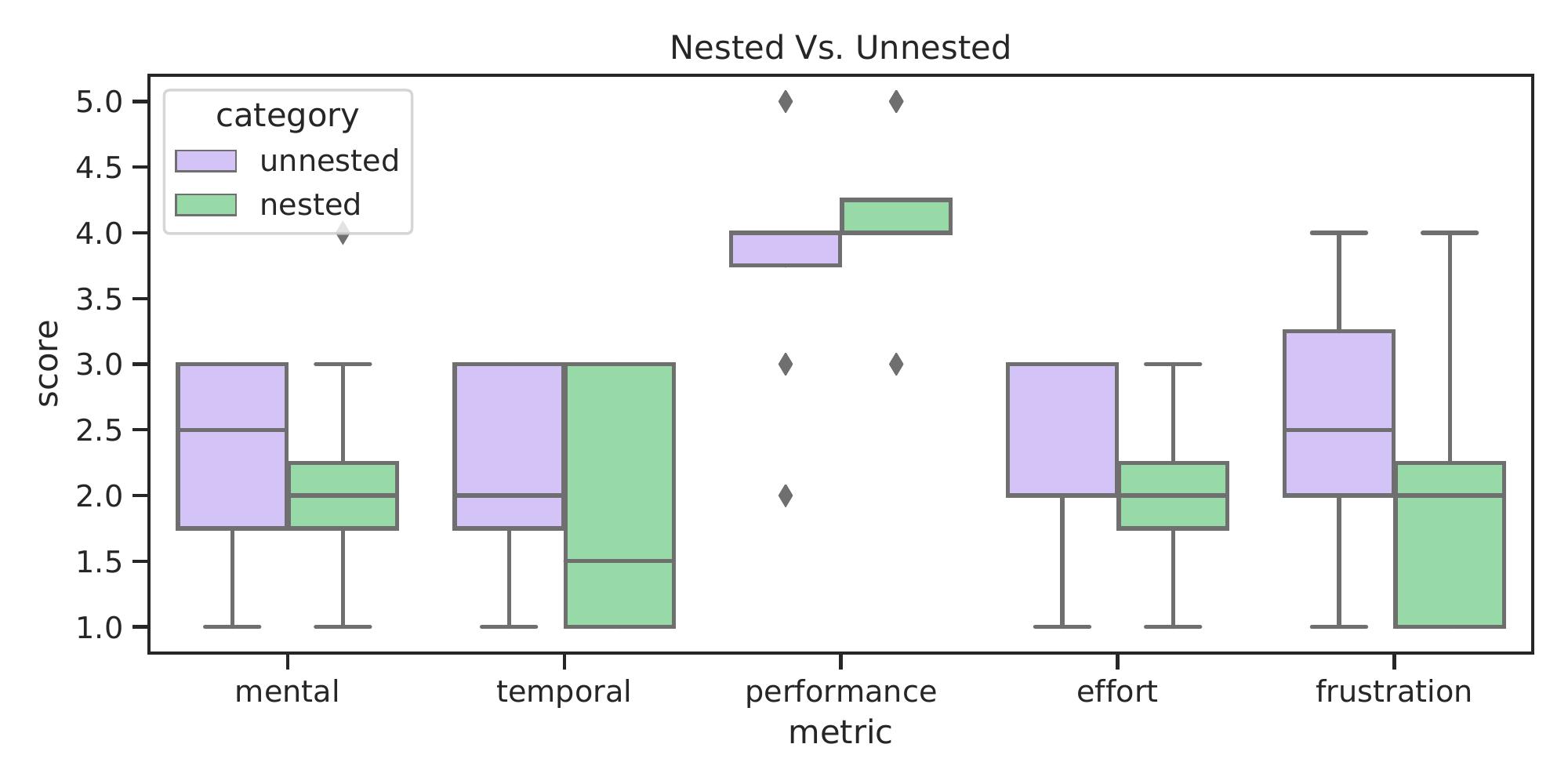}
  \caption{NASA TLX results from questionnaire on having users specify functions by nesting them vs. not. }
  \label{fig:tasklet_3}
\end{figure}

\subsubsection{Nesting Function Specification} 
\VASH allows a function's specification to be nested in another.  We created a test where users created a skill to buy ingredients for a recipe.  They need to create a ``buy'' function which is then used on a ``recipe'' function which applies ``buy'' on a list of ingredients. We asked them to do it by (1) using the nested specification method, and (2) first defining the ``buy'' function then the ``recipe'' function. 
Figure~\ref{fig:tasklet_3} shows the  NASA-TLX~\cite{hart2006nasa} survey results from this study. Across all metrics, the results point to users preferring the nested specification method. (Note that a low number is better for all but the performance question).


\subsection{Real Scenarios Evaluation}
In our final  experiment, we want to get users' feedback of using the system in real life tasks drawn from our first user study, using websites they are familiar with: Walmart, a recipe website, a stock website, and a weather website. This end-to-end test also demonstrates that \VASH is a fully functional system. 



Each task involves the user defining a \VASH skill, and then invoking it to see the result. We evaluated the following real-world scenarios, which we chose based on the need-finding experiment:
\begin{enumerate}
\item \textit{Calculate the average of temperatures}.
The user creates a program that goes to weather.gov, enters their zip code, calculates the average high temperature for the week, and returns that value.  This example exercises the multi-selection and aggregation function. 

\item \textit{Add items to an online shopping cart}.
Here, the user has a shopping list of items that they entered, and they need to write a program that add them all to a shopping cart on everlane.com.
This scenario requires user input, copy and paste, and iteration.

\item \textit{Notify when stock prices dip}.
The user creates a skill on zacks.com to receive a notification when a stock quote goes under a fixed price.  The skill is then triggered each day at a certain time.  This tests the conditional and timer functions of the system.

\item \textit{Add ingredients from a website to a shopping cart}.
This task is similar to the task in Fig.~\ref{fig:feature-graphic}.  
The user visits a cooking website, acouplecooks.com,
and uses their keyboard to copy an ingredient on a website to their clipboard.  They then go to walmart.com, paste the ingredient in the search bar, click on the first item, then add it to their cart.  They go back to the cookie recipe and add the rest of the items to their cart iteratively using the function created.  This tests users' understanding of creating skills using different websites and defining a function within a function.
\end{enumerate}
\begin{figure}
  \includegraphics[width=\linewidth]{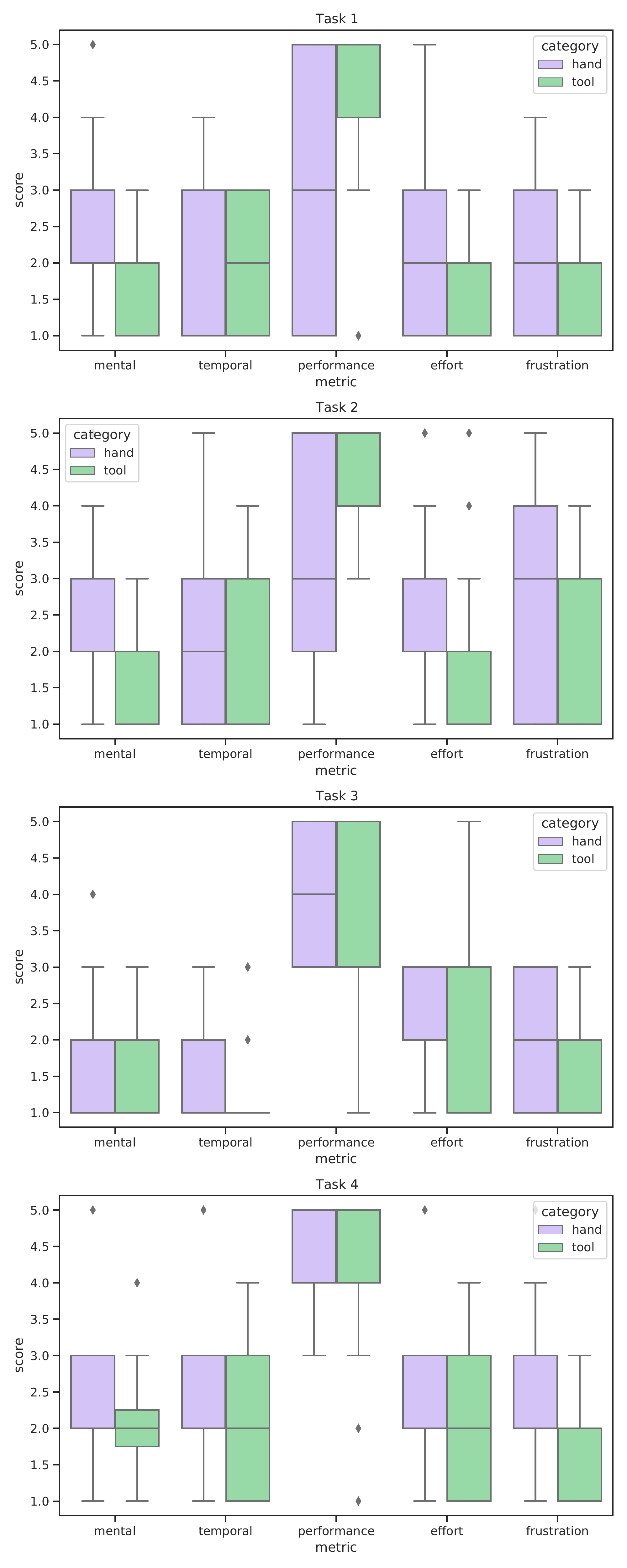}
  \caption{For each of the tasks in the real world scenarios we use the NASA-TLX score to evaluate the perceived workload of complete the task using \VASH compared with doing it by hand. Lower scores are better in all categories, except for performance, where higher values are better.}
  \label{fig:nasa_real_world}
\end{figure}

We conducted this study as an interactive user test (live over video-conferencing) using the same participants as the design decision study described above.
Users first complete a warm-up task of recording a simple function to familiarize with \VASH. Then they are asked to complete each task manually and on \VASH following a predefined script. Whether each user completes the task manually or using \VASH first is randomized. After each task, we ask the user to complete the NASA-TLX survey, comparing the system to performing the task by hand. 

All users were able to install \VASH on their Chrome browser and complete the tasks successfully. Fig.~\ref{fig:nasa_real_world} shows the result of the NASA-TLX survey, aggregated across all tasks. We first observe that there is no statistically significant difference across all five metrics between completing the tasks by hand and completing it using \VASH. This suggests that recording a skill with \VASH is no harder than performing the same task by hand, while providing the immediate value of being able to invoke the skill repeatedly. 
This is a promising result that shows automation can be practical for end-users.

Users also complete a Likert-scale evaluation on the whole system. This survey shows how users would perceive its value with real-world scenarios. The results are shown in Fig.~\ref{figure:likert-scale} as ``Exp. B''. 73\% of the users find the system easy to learn, but only 46\% find it easy to use, probably due to the complexity of the tested tasks. 47\% of the users find PBD is useful, and 73\% find the multimodal interface useful. Overall, a large majority of the users (80\%) find \VASH useful, and 67\% are satisfied with it after testing. Once again, a majority of users (53\%) agree they would personally use \VASH. 

\subsubsection{Qualitative Feedback}
During the user test, we also collected qualitative feedback from the participants.
A user that was not able to program before said, ``when you're raised on sci-fi movies, the thought of a system that can learn what you need by voice is incredibly appealing.'' A user saw the system as being very helpful in repeating common tasks to accomplish her job, ``for me as a data person especially, during the COVID-19 crisis when local governments are behind on data standards, I've found the lack of such tool exhausting. The level of manual data entry required to achieve my basic analysis goals is often more than I can make time for, and one day that I fail to check is data that may be permanently lost. I love the idea of being able to program that cleanly, with my voice. I love that it can intelligently extract numbers from characters and perform basic operations, and run just by speaking.'' 


%% file: limitations.tex
\section{Limitations and Future Work}

\subsection{Brittle Selectors}
Our system records objects a user is referring to when performing an action using their CSS selectors. Hence, our system can fail whenever the CSS selectors on the web page change. This can be due to a change in a website's layout, recompiling dynamic CSS modules, or automatically generated CSS classes. Additionally, our system cannot select a portion of an element, and must always select the full element. Future work should investigate the use of alternative selection strategies, such as combining clicks with natural language specifications~\cite{li2018appinite}, or selection based on computer vision~\cite{Sereshkeh_2020}.

\subsection{Robustness of Execution with Puppeteer}
In addition to the limitations of CSS selectors, another source of potential failures while executing a \WebTalk is due to timing, and the amount of dynamic content. For example, inserting variables using Puppeteer in Google's search bar was very inconsistent due to the input's extensive use of JavaScript for auto-complete. To allow pages to react to the user input, we rate-limited the calls to the Puppeteer API, but that made the system quite slow (about as slow as executing by hand). In the future, we should introduce a mechanism to detect dynamically when the page is changing and when it is stable and ready for the next action.

\subsection{Natural Language and Voice Recognition}
Our current prototype uses a template-based algorithm to understand the user's commands in natural language. As a result, users have less freedom to alter the wording of their utterances slightly, requiring them to memorize commands. Users have reported in the user study that remembering commands was non-trivial. Future work should investigate the use of neural semantic parsing to translate from natural language to \WebTalk statements directly~\cite{geniepldi19}. Using a neural semantic parser would also allow users to compound multiple statements or clauses in a single sentence, for example combine a timer and a conditional.

\subsection{Error Correction and Editability}
Once a user defines a function by demonstration, the program is not editable either via demonstration or through direct modification of the \WebTalk code. This is particularly problematic because for conditionals, we can only demonstrate one side of the conditional branch. In the future users will be able to refine their created functions, either by recording additional traces that the system would merge, or by interactively editing the generated code (perhaps in natural language). Interactive refinement is also a prerequisite for building larger skills, which need to be debugged.

\subsection{Authentication \& Privacy}
As the browser profiles used for recording and execution are entirely separate, our design requires the execution browser to replicate the entirety of the profile (cookies, local storage, certificates, saved passwords, etc.). In \VASH, this is acceptable because both browsers are run on the user's machine. At the same time, giving the execution browser access to all the user's cookies can be a significant privacy loss, if  the execution browser runs on a separate server, which is desirable for availability and for performance purposes. 

\subsection{Authentication, Fraud, and Spam}
\VASH can fail to work well with websites such as GMail and Facebook that use systems to detect fraudulent logins and bots.  Such anti-fraud systems can detect the use of automated browsing APIs, and can detect input that is driven by a program as opposed to a user with keyboard and mouse. Various techniques have been proposed to subvert these detection mechanisms~\cite{puppeteer-extra-plugin-stealth} but as the subversion improves, so does the anti-fraud detection.

%% file: conclusion.tex
\section{Conclusion}
Virtual assistants are changing the way we interact with computers. Along with this, we need to empower individuals to build programs for virtual assistants, instead of having to reply on only applications built by large companies. 

This paper proposes \VASH, a multimodal system that lets users create their own web skills using program by demonstration. \VASH translates voice, keyboard, and mouse inputs into a program in WebTalk, a language we created for this purpose. 
\VASH supports control constructs: function calls, conditionals, and iterations, which are all found to be essential for automating tasks our users in our survey want.  We find \VASH to be expressive enough to implement 81\% of user-proposed skills. 
We also evaluate several main design choices, showing them to be effective. We find that \VASH is easy to learn, and that a majority of the users in our study want to use our system. 

In summary, \VASH is an easy-to-learn system that lets consumers create useful virtual assistant web-based skills that require sophisticated control constructs.